\documentclass[vecphys]{svmult}

\usepackage{makeidx}         
\usepackage{multicol}        
\usepackage[bottom]{footmisc}

\makeindex             

\def\square{\kern1pt\vbox{\hrule height 1.2pt\hbox{\vrule width 1.2pt\hskip 3pt
   \vbox{\vskip 6pt}\hskip 3pt\vrule width 0.6pt}\hrule height 0.6pt}\kern1pt}

\begin{document}

\title*{Avoiding Dark Energy with 1/R Modifications of Gravity}
\titlerunning{1/R Modifications of Gravity}

\author{R. P. Woodard}

\institute{Department of Physics, University of Florida, Gainesville, FL
32611-8440, USA
\texttt{woodard@phys.ufl.edu}}

\maketitle

\section{Introduction}
\label{sec:1}

The case for alternate gravity is easily made. The best that can be done from 
observing cosmic motions is to infer the metric $g_{\mu\nu}$ in some 
coordinate system. From this one can reconstruct the Einstein tensor and then 
ask whether or not general relativity predicts it in terms of the observed 
sources of stress-energy,
\begin{equation}
\Bigl(R_{\mu\nu} - \frac12 g_{\mu\nu} R\Bigr)_{\rm rec} = 8 \pi G \Bigl(
T_{\mu\nu}\Bigr)_{\rm obs} \; ?
\end{equation}
One way of explaining any disagreement is by positing the existence of an 
unobserved, ``dark'' component of the stress-energy tensor,
\begin{equation}
\Bigl(T_{\mu\nu}\Bigr)_{\rm dark} \equiv \frac1{8\pi G} \Bigl(R_{\mu\nu}
- \frac12 g_{\mu\nu} R\Bigr)_{\rm rec} - \Bigl(T_{\mu\nu}\Bigr)_{\rm obs} \; .
\end{equation}
This always works, but recent observations make it seem epicyclic. 

The theory of nucleosynthesis implies that no more than about 4\% of the 
energy density currently required to make general relativity agree with all 
observations can consist of any material with which we are presently familiar 
\cite{BBN} --- and only a fraction of this 4\% is observed. Just to 
make general relativity agree with the observed motions of galaxies and 
galactic clusters we must posit that {\it six times} the mass of ordinary 
matter comes in the form of nonbaryonic, cold dark matter \cite{CDM}. 
Although there are some plausible candidates for what this might be, no 
Earth-bound laboratory has yet succeeded in detecting it. 

I belong to the minority of physicists who feel that this factor of six 
already strains
credulity. Easing that strain is what led Milgrom to propose MOND \cite{MM}, 
which can be viewed as a phenomenological modification of gravity in the 
regime of very small accelerations. There is an impressive amount of
observational data in favor of this modification \cite{SM} --- although see
\cite{GSKVK}. Bekenstein has recently constructed a fully relativistic field 
theory \cite{JDB} which reproduces MOND, and a preliminary analysis of the 
resulting cosmology works better than many experts thought possible 
\cite{SMFB}.

However, the worst problem for conventional gravity comes on the largest 
scales. To make general relativity agree with the Hubble plots of distant 
Type Ia supernovae \cite{SNCP,SNST,SNLS}, with the power spectrum of 
anisotropies in the cosmic
microwave background \cite{CMB} and with large scale structure surveys 
\cite{LSS}, one must accept an additional component of ``dark energy'' that 
is about {\it eighteen times} larger than that of ordinary matter. This 
would mean that 96\% of the current universe's energy exists in forms which
have so far only been detected gravitationally! Even people who believe
passionately in dark matter (and hence accept the factor of six) find this
factor of $6 \!+\! 18 \!=\! 24$ difficult to swallow. That is why there has 
been so much recent interest in modifying gravity to make it predict 
observed cosmic phenomena without the need for dark energy, and sometimes 
even without the need for dark matter.

I want to stress that the issue is one of plausibility. There is no
problem inventing field theories which give the required amount of dark
energy. The simplest way of doing it is with a minimally coupled scalar
\cite{CW,PR},
\begin{equation}
\mathcal{L} = -\frac12 \partial_{\mu} \varphi \partial_{\nu} g^{\mu\nu}
\sqrt{-g} - V(\varphi) \sqrt{-g} \; . \label{quint}
\end{equation}
The usual procedure is to begin with a scalar potential $V(\varphi)$ and
work out the cosmology, but it is easy to start with whatever cosmological
evolution is desired and {\it construct} the potential which would support
it. I will go through the construction here, both to make the point and so
that it can be used later.

On the largest scales the geometry of the universe can be described in terms 
of a single function of time known as the scale factor $a(t)$,
\begin{equation}
ds^2 = -dt^2 + a^2(t) d\vec{x} \cdot d\vec{x} \; .
\end{equation}
The logarithmic time derivative of this quantity gives the Hubble parameter,
\begin{equation}
H(t) \equiv \frac{\dot{a}}{a} \; .
\end{equation}
If we specialize to a solution $\varphi_0(t)$ of the scalar field equations 
which depends only upon time, the two nontrivial Einstein equations are,
\begin{eqnarray}
3 H^2 & = & 8 \pi G \Bigl(\frac12 \dot{\varphi}_0^2 + V(\varphi_0)\Bigr) \; ,
\label{E1} \\
-2 \dot{H} - 3 H^2 & = & 8 \pi G \Bigl(\frac12 \dot{\varphi}_0^2 - 
V(\varphi_0)\Bigr) \; . \label{E2}
\end{eqnarray}
Let us assume $a(t)$ is known as an explicit function of time, and construct
$\varphi_0(t)$ and $V(\varphi)$. By adding (\ref{E1}) and (\ref{E2}) we obtain,
\begin{equation}
-2 \dot{H} = 8 \pi G \dot{\varphi}_0^2 \; . \label{twoeqns}
\end{equation}
The weak energy condition implies $\dot{H}(t) \leq 0$ so we can take the
square root and integrate to solve for $\varphi_0(t)$,
\begin{equation}
\varphi_0(t) = \varphi_I \pm \int_{t_I}^t dt' \sqrt{\frac{-2 \dot{H}(t')}{
8 \pi G}} \; . \label{phi}
\end{equation}
One can choose $\varphi_I$ and the sign freely.

Because the integrand in (\ref{phi}) is always positive, the function
$\varphi_0(t)$ is monotonic. This means we can invert to solve for
time as a function of $\varphi_0$. Let us call the inverse function 
$T(\varphi)$,
\begin{equation}
\psi = \varphi_0\Bigl(T(\psi)\Bigr) \; . \label{inv}
\end{equation} 
By subtracting (\ref{E2}) from (\ref{E1}) we obtain a relation for the
scalar potential as a function of time,
\begin{equation}
V = \frac1{8\pi G} \Bigl( \dot{H}(t) + 3 H^2(t)\Bigr) \; .
\end{equation}
The potential is determined as a function of the scalar by substituting the 
inverse function (\ref{inv}),
\begin{equation}
V(\varphi) = \frac1{8\pi G} \Biggl\{ \dot{H}\Bigl(T(\varphi)\Bigr) 
+ 3 H^2\Bigl(T(\varphi)\Bigr) \Biggr\} \; .
\end{equation}

This construction gives a scalar which supports any evolution $a(t)$ (with
$\dot{H}(t) < 0$) all by itself. Should you wish to include some other, 
known component of the stress-energy, simply add the energy density 
and pressure of this component to the Einstein equations,
\begin{eqnarray}
3 H^2 & = & 8 \pi G \Bigl(\frac12 \dot{\varphi}_0^2 + V(\varphi_0) + \rho_{\rm
known}\Bigr) \; , \\
-2 \dot{H} - 3 H^2 & = & 8 \pi G \Bigl(\frac12 \dot{\varphi}_0^2 - 
V(\varphi_0) + p_{\rm known}\Bigr) \; .
\end{eqnarray}
Provided $\rho_{\rm known}$ and $p_{\rm known}$ are known functions of either
time or the scale factor, the construction goes through as before.\footnote{
This construction seems to be due to Ratra and Peebles \cite{PR}. Recent
examples of its use include \cite{TW2,SRSS,NO0}.}

Using this method one can devise a new field $\varphi(x)$ which will
support {\it any} cosmology with $\dot{H}(t) < 0$. However, the introduction 
of such a ``quintessence'' field raises a number of questions:
\begin{enumerate}
\item{Where does $\varphi$ reside in fundamental theory?}
\item{Why can't $\varphi$ couple to fields other than the metric? And if
it does couple to other fields, why haven't we detected its influence in 
Earth-bound laboratories?}
\item{Why did $\varphi$ come to dominate the stress-energy of the universe
so recently in cosmological time?}
\item{Why is the $\varphi$ field so homogeneous?}
\end{enumerate}
When a phenomenological fix raises more questions than it answers people
are naturally drawn to investigate other fixes. One possibility is that 
general relativity is not the correct theory of gravity on cosmological 
scales. 

In this talk I shall review gravitational Lagrangians of the form,
\begin{equation}
\mathcal{L} = \frac1{16 \pi G} \Bigl(R + \Delta R[g]\Bigr) \sqrt{-g} \; ,
\label{ansatz}
\end{equation}
where $\Delta R[g]$ is some local scalar constructed from the curvature
tensor and possibly its covariant derivatives. Examples of such scalars
are,
\begin{equation}
\frac1{\mu^2} R^{\alpha\beta} R_{\alpha\beta} \qquad , \qquad \frac1{\mu^4}
g^{\mu\nu} R_{,\mu} R_{,\nu} \qquad , \qquad \mu^2 \sin\Bigl(\frac1{\mu^4}
R^{\alpha\beta\rho\sigma} R_{\alpha\beta\rho\sigma}\Bigr) \; .
\end{equation}
I begin by reviewing a powerful no-go theorem which pervades and constrains
fundamental theory so completely that most people assume its consequence
without thinking. This is the theorem of Ostrogradski \cite{MO}, who 
essentially showed why Newton was right to suppose that the laws of physics 
involve no more than two time derivatives of the fundamental dynamical 
variables. The key consequence for our purposes is that the only viable
form for the functional $\Delta R[g]$ in (\ref{ansatz}) is an algebraic 
function of the undifferentiated Ricci scalar,
\begin{equation}
\Delta R[g] = f(R) \; .
\end{equation}
I review the Ostrogradski result in section 2, and hopefully immunize you
against some common misconceptions about it in section 3. In section 4 I 
explain why $f(R)$ theories do not contradict Ostrogradski's result. I also
demonstrate that, in the absence of matter, $f(R)$ theories are equivalent 
to ordinary gravity, with $f(R) = 0$, plus a minimally coupled scalar of 
the form (\ref{quint}). Then I use the construction given above to show how 
one can choose $f(R)$ to enforce an arbitrary cosmology. This establishes 
that an $f(R)$ can be found to support any desired cosmology. In section 5 
I discuss problems associated with the particular choice function $f(R) = 
-\frac{\mu^4}{R}$. Section 6 presents conclusions.

\section{The Theorem of Ostrogradski}
\label{sec:2}

Ostrogradski's result is that there is a linear instability in the Hamiltonians
associated with Lagrangians which depend upon more than one time derivative in 
such a way that the dependence cannot be eliminated by partial integration
\cite{MO}. The result is so general that I can simplify the discussion by
presenting it in the context of a single, one dimensional point particle whose
position as a function of time is $q(t)$. First I will review the way the
Hamiltonian is constructed for the usual case in which the Lagrangian involves
no higher than first time derivatives. Then I present Ostrogradski's 
construction for the case in which the Lagrangian involves second time
derivatives. And the section closes with the generalization to $N$ time
derivatives.

In the usual case of $L = L(q,\dot{q})$, the Euler-Lagrange equation is,
\begin{equation}
\frac{\partial L}{\partial q} - \frac{d}{dt} \frac{\partial L}{\partial
\dot{q}} = 0 \; . \label{ELE1}
\end{equation}
The assumption that $\frac{\partial L}{\partial \dot{q}}$ depends upon 
$\dot{q}$ is known as {\it nondegeneracy}. If the Lagrangian is nondegenerate
we can write (\ref{ELE1}) in the form Newton assumed so long ago for the laws 
of physics,
\begin{equation}
\ddot{q} = \mathcal{F}(q,\dot{q}) \qquad \Longrightarrow \qquad q(t) =
\mathcal{Q}(t,q_0,\dot{q}_0) \; . \label{newt}
\end{equation}
From this form it is apparent that solutions depend upon two pieces of initial
value data: $q_0 = q(0)$ and $\dot{q}_0 = \dot{q}(0)$. 

The fact that solutions require two pieces of initial value data means that
there must be two canonical coordinates, $Q$ and $P$. They are traditionally
taken to be,
\begin{equation}
Q \equiv q \qquad {\rm and} \qquad P \equiv \frac{\partial L}{\partial \dot{q}}
\; . \label{ctrans}
\end{equation}
The assumption of nondegeneracy is that we can invert the phase space
transformation (\ref{ctrans}) to solve for $\dot{q}$ in terms of $Q$ and $P$. 
That is, there exists a function $v(Q,P)$ such that,
\begin{equation}
\frac{\partial L}{\partial \dot{q}} \Biggl\vert_{q = Q \atop \dot{q} = v}
= P \; . \label{invct}
\end{equation}

The canonical Hamiltonian is obtained by Legendre transforming on $\dot{q}$,
\begin{eqnarray}
H(Q,P) & \equiv & P \dot{q} - L \; , \\
& = & P v(Q,P) - L\Bigl(Q,v(Q,P)\Bigr) \; .
\end{eqnarray}
It is easy to check that the canonical evolution equations reproduce the 
inverse phase space transformation (\ref{invct}) and the Euler-Lagrange 
equation (\ref{ELE1}),
\begin{eqnarray}
\dot{Q} & \equiv & \frac{\partial H}{\partial P} = v + P \frac{\partial v}{
\partial P} - \frac{\partial L}{\partial \dot{q}} \frac{\partial v}{\partial P}
= v \; , \\
\dot{P} & \equiv & -\frac{\partial H}{\partial Q} = -P \frac{\partial v}{
\partial Q} + \frac{\partial L}{\partial q} + \frac{\partial L}{\partial
\dot{q}} \frac{\partial v}{\partial P} = \frac{\partial L}{\partial q} \; .
\end{eqnarray}
This is what we mean by the statement, ``the Hamiltonian generates time
evolution.'' When the Lagrangian has no explicit time dependence, $H$ is also
the associated conserved quantity. Hence it is ``the'' energy by anyone's
definition, of course up to canonical transformation.

Now consider a system whose Lagrangian $L(q,\dot{q},\ddot{q})$ depends 
nonde\-gen\-er\-ate\-ly upon $\ddot{q}$. The Euler-Lagrange equation is,
\begin{equation}
\frac{\partial L}{\partial q} - \frac{d}{dt} \frac{\partial L}{\partial 
\dot{q}} + \frac{d^2}{dt^2} \frac{\partial L}{\partial \ddot{q}} = 0 \; .
\label{ELE2}
\end{equation}
Non-degeneracy implies that $\frac{\partial L}{\partial \ddot{q}}$ depends
upon $\ddot{q}$, in which case we can cast (\ref{ELE2}) in a form radically
different from Newton's,
\begin{equation}
q^{(4)} = \mathcal{F}(q,\dot{q},\ddot{q},q^{(3)}) \qquad \Longrightarrow \qquad
q(t) = \mathcal{Q}(t,q_0,\dot{q}_0,\ddot{q}_0,q^{(3)}_0) \; .
\end{equation}

Because solutions now depend upon four pieces of initial value data there must
be four canonical coordinates. Ostrogradski's choices for these are,
\begin{eqnarray}
Q_1 \equiv q \qquad & , & \qquad P_1 \equiv \frac{\partial L}{\partial \dot{q}}
- \frac{d}{dt} \frac{\partial L}{\partial \ddot{q}} \; , \label{ct1} \\
Q_2 \equiv \dot{q} \qquad & , & \qquad P_2 \equiv \frac{\partial L}{\partial 
\ddot{q}} \; . \label{ct2}
\end{eqnarray}
The assumption of nondegeneracy is that we can invert the phase space
transformation (\ref{ct1}-\ref{ct2}) to solve for $\ddot{q}$ in terms of
$Q_1$, $Q_2$ and $P_2$. That is, there exists a function $a(Q_1,Q_2,P_2)$
such that,
\begin{equation}
\frac{\partial L}{\partial \ddot{q}} \Biggl\vert_{{q = Q_1 \atop \dot{q} =
Q_2} \atop \ddot{q} = a} = P_2 \; . \label{invct2}
\end{equation}
Note that one only needs the function $a(Q_1,Q_2,P_2)$ to depend upon {\it
three} canonical coordinates --- and not all four --- because 
$L(q,\dot{q},\ddot{q})$ only depends upon three configuration space 
coordinates. This simple fact has great consequence.

Ostrogradski's Hamiltonian is obtained by Legendre transforming, just as in the
first derivative case, but now on $\dot{q} = q^{(1)}$ and $\ddot{q} = q^{(2)}$,
\begin{eqnarray}
\lefteqn{H(Q_1,Q_2,P_1,P_2) \equiv \sum_{i=1}^2 P_i q^{(i)} - L \; , } \\
& & = P_1 Q_2 + P_2 a(Q_1,Q_2,P_2) - L\Bigl(Q_1,Q_2,a(Q_1,Q_2,P_2)\Bigr) \; .
\label{Host}
\end{eqnarray}
The time evolution equations are just those suggested by the notation,
\begin{equation}
\dot{Q_i} \equiv \frac{\partial H}{\partial P_i} \qquad {\rm and} \qquad 
\dot{P}_i \equiv - \frac{\partial H}{\partial Q_i} \; .
\end{equation}
Let's check that they generate time evolution. The evolution equation for
$Q_1$,
\begin{equation}
\dot{Q}_1 = \frac{\partial H}{\partial P_1} = Q_2 \; ,
\end{equation}
reproduces the phase space transformation $\dot{q} = Q_2$ in (\ref{ct2}).
The evolution equation for $Q_2$,
\begin{equation}
\dot{Q}_2 = \frac{\partial H}{\partial P_2} = a + P_2 \frac{\partial a}{
\partial P_2} - \frac{\partial L}{\partial \ddot{q}} \frac{\partial a}{\partial
P_2} = a \; ,
\end{equation}
reproduces (\ref{invct2}). The evolution equation for $P_2$,
\begin{equation}
\dot{P}_2 = -\frac{\partial H}{\partial Q_2} = -P_1 - P_2 \frac{\partial a}{
\partial Q_2} + \frac{\partial L}{\partial \dot{q}} + \frac{\partial L}{
\partial \ddot{q}} \frac{\partial a}{\partial Q_2} = -P_1 + \frac{\partial L}{
\partial \dot{q}} \; ,
\end{equation}
reproduces the phase space transformation $P_1 = \frac{\partial L}{\partial
\dot{q}} - \frac{d}{dt} \frac{\partial L}{\partial \ddot{q}}$ (\ref{ct1}). And 
the evolution equation for $P_1$,
\begin{equation}
\dot{P}_1 = -\frac{\partial H}{\partial Q_1} = -P_2 \frac{\partial a}{\partial 
Q_1} + \frac{\partial L}{\partial q} + \frac{\partial L}{\partial \ddot{q}}
\frac{\partial a}{\partial Q_1} = \frac{\partial L}{\partial q} \; ,
\end{equation}
reproduces the Euler-Lagrange equation (\ref{ELE2}). So Ostrogradski's system
really does generate time evolution. When the Lagrangian contains no explicit
dependence upon time it is also the conserved Noether current. By anyone's
definition, it is therefore ``the'' energy, again up to canonical 
transformation.

There is one, overwhelmingly bad thing about Ostrogradski's Hamiltonian 
(\ref{Host}): it is {\it linear} in the canonical momentum $P_1$. This means 
that no system of this form can be stable. In fact, there is not even any 
barrier to decay. Note also the power and generality of the result. It applies
to {\it every} Lagrangian $L(q,\dot{q},\ddot{q})$ which depends 
nondegenerately upon $\ddot{q}$, independent of the details. The only 
assumption is nondegeneracy, and that simply means one cannot eliminate 
$\ddot{q}$ by partial integration. This is why Newton was right to assume 
the laws of physics take the form (\ref{newt}) when expressed in terms of 
fundamental dynamical variables.

Adding more higher derivatives just makes the situation worse. Consider a 
Lagrangian $L\left(q,\dot{q},\dots,q^{(N)}\right)$ which depends upon the 
first $N$ derivatives of $q(t)$. If this Lagrangian depends nondegenerately
upon $q^{(N)}$ then the Euler-Lagrange equation,
\begin{equation}
\sum_{i=0}^N \left(-{d \over dt}\right)^i {\partial L \over \partial q^{(i)}}
= 0 \; , \label{ELEN}
\end{equation}
contains $q^{(2N)}$. Hence the canonical phase space must have $2N$ 
coordinates. Ostrogradski's choices for them are,
\begin{equation}
Q_i \equiv q^{(i-1)} \qquad {\rm and} \qquad P_i \equiv \sum_{j=i}^N \Bigl(-
\frac{d}{dt}\Bigr)^{j-i} \frac{\partial L}{\partial q^{(j)}} \; .
\end{equation}
Non-degeneracy means we can solve for $q^{(N)}$ in terms of $P_N$ and the
$Q_i$'s. That is, there exists a function $\mathcal{A}(Q_1,\ldots,Q_N,P_N)$
such that,
\begin{equation}
\frac{\partial L}{\partial q^{(N)}} \Biggl\vert_{q^{(i-1)} = Q_i \atop
q^{(N)} = \mathcal{A}} = P_N \; . \label{nondeg}
\end{equation}

For general $N$ Ostrogradski's Hamiltonian takes the form,
\begin{eqnarray}
H & \equiv & \sum_{i=1}^N P_i q^{(i)} - L \; , \\
& = & P_1 Q_2 + P_2 Q_3 + \cdots + P_{N-1} Q_N + P_N \mathcal{A} -
L\Bigl(Q_1,\ldots,Q_N,\mathcal{A}\Bigr) \; . \label{HN}
\end{eqnarray}
It is simple to check that the evolution equations,
\begin{equation}
\dot{Q}_i \equiv \frac{\partial H}{\partial P_i} \qquad {\rm and} \qquad 
\dot{P}_i \equiv -\frac{\partial H}{\partial Q_i} \; ,
\end{equation}
again reproduce the canonical transformations and the Euler-Lagrange equation.
So (\ref{HN}) generates time evolution. Similarly, it is Noether current for 
the case where the Lagrangian contains no explicit time dependence. So there
is little alternative to regarding (\ref{HN}) as ``the'' energy, again up to
canonical transformation.

One can see from (\ref{HN}) that the Hamiltonian is linear in $P_1, P_2, \ldots
P_{N-1}$. Only with respect to $P_N$ might it be bounded from below. Hence
the Hamiltonian is necessarily unstable over half the classical phase space
for large $N$!

\section{Common Misconceptions}
\label{sec:3}

The no-go theorem I have just reviewed ought to come as no surprise. It 
explains why Newton was right to expect that physical laws take the form
of second order differential equations when expressed in terms of fundamental
dynamical variables.\footnote{The caveat is there because one can always get 
higher order equations by solving for some of the fundamental variables.} 
Every fundamental system we have discovered since Newton's day has had this 
form. The bizarre, dubious thing would be if Newton had blundered upon a tiny 
subset of possible physical laws, and all our probing over the course of the 
next three centuries had never revealed the vastly richer possibilities. 
However --- {\it deep sigh} --- particle theorists don't like being told 
something is impossible, and a definitive no-go theorem such as that of 
Ostrogradski provokes them to tortuous flights of evasion. I ought to know, I 
get called upon to referee the resulting papers often enough! No one has so 
far found a way around Ostrogradski's theorem. I won't attempt to prove that 
no one ever will, but let me use this section to run through some of the 
misconceptions which have been in back of attempted evasions.

To fix ideas it will be convenient to consider a higher derivative 
generalization of the harmonic oscillator,
\begin{equation}
\mathcal{L} = -\frac{g m}{2 \omega^2} \ddot{q}^2 + \frac{m}2 \dot{q}^2 -
\frac{m\omega^2}2 q^2 \; . \label{HDO}
\end{equation}
Here $m$ is the particle mass, $\omega$ is a frequency and $g$ is a small
positive pure number we can think of as a coupling constant. The Euler-Lagrange 
equation,
\begin{equation}
-m \Bigl( \frac{g}{\omega^2} q^{(4)} + \ddot{q} + \omega^2 q\Bigr) = 0 \; ,
\label{HDE}
\end{equation}
has the general solution,
\begin{equation}
q(t) = A_+ \cos(k_+ t) + B_+ \sin(k_+ t) + A_- \cos(k_- t) + B_- \sin(k_- t) 
\; . \label{gensol}
\end{equation}
Here the two frequencies are,
\begin{equation}
k_{\pm} \equiv \omega \sqrt{ \frac{1 \mp \sqrt{1 \!-\! 4 g}}{2 g} } \; ,
\end{equation}
and the initial value constants are,
\begin{eqnarray}
A_+ = \frac{k_-^2 q_0 \!+\! \ddot{q}_0}{k_-^2 \!-\! k_+^2} \qquad & , & \qquad
B_+ = \frac{k_-^2 \dot{q}_0 \!+\! q^{(3)}_0}{k_+ (k_-^2 \!-\! k_+^2)} \; , \\
A_- = \frac{k_+^2 q_0 \!+\! \ddot{q}_0}{k_+^2 \!-\! k_-^2} \qquad & , & \qquad
B_- = \frac{k_+^2 \dot{q}_0 \!+\! q^{(3)}_0}{k_- (k_+^2 \!-\! k_-^2)} \; .
\end{eqnarray}
The conjugate momenta are,
\begin{eqnarray}
P_1 = m \dot{q} + \frac{g m}{\omega^2} q^{(3)} \qquad & \Leftrightarrow & 
\qquad q^{(3)} = \frac{\omega^2 P_1 \!-\! m \omega^2 Q_2}{g m} \; , \\
P_2 = - \frac{g m}{\omega^2} \ddot{q} \qquad & \Leftrightarrow & 
\qquad \ddot{q} = -\frac{\omega^2 P_2}{g m} \; .
\end{eqnarray}
The Hamiltonian can be expressed in terms of canonical variables,
configuration space variables or initial value constants,
\begin{eqnarray}
H & = & P_1 Q_2 - \frac{\omega^2}{2 g m} P_2^2 - \frac{m}2 Q_2^2 + \frac{m
\omega^2}2 Q_1^2 \; , \label{H1} \\
& = & \frac{g m}{\omega^2} \dot{q} q^{(3)} - \frac{g m}{2 \omega^2}
\ddot{q}^2 + \frac{m}2 \dot{q}^2 + \frac{m \omega^2}2 q^2 \; , \label{H2} \\
& = & \frac{m}2 \sqrt{1 \!-\! 4 g} \, k_+^2 (A_+^2 \!+\! B_+^2) - 
\frac{m}2 \sqrt{1 \!-\! 4 g} \, k_-^2 (A_-^2 \!+\! B_-^2) \; . \label{H3}
\end{eqnarray}
The last form makes it clear that the ``$+$'' modes carry positive energy
whereas the ``$-$'' modes carry negative energy.

\subsection{Nature of the Instability}
\label{sub:3.1}

It's important to understand both how the Ostrogradskian instability manifests
and what is physically wrong with a theory which shows this instability. 
Because the Ostrogradskian Hamiltonian (\ref{HN}) is not bounded below with 
respect to more than one of its conjugate momenta, one sees that the problem 
is not reaching arbitrarily negative energies by setting the dynamical
variable to some {\it constant value}. Rather it is reaching arbitrarily
negative energies by making the dynamical variable have a certain {\it time 
dependence}. People sometimes mistakenly believe they have found a higher
derivative system which is stable when all they have checked is that the
Hamiltonian is bounded from below for constant field configurations. For
example, from expression (\ref{H2}) we see that our higher derivative
oscillator energy is bounded below by zero for $q(t) = {\rm const}$! Negative 
energies are achieved by making $\ddot{q}$ large and/or making $q^{(3)}$ 
large while keeping $\dot{q} \!+\! g q^{(3)}/\omega^2$ fixed.

Another crucial point is that the same dynamical variable typically carries
both positive and negative energy degrees of freedom in a higher derivative
theory. For our higher derivative oscillator this is apparent from expression
(\ref{gensol}) which shows that $q(t)$ involves both the positive energy
degrees of freedom, $A_+$ and $B_+$, and the negative energy ones, $A_-$ and
$B_-$. And note from expression (\ref{H3}) that I really mean positive and 
negative {\rm energy}, not just positive and negative frequency, which is the
usual case in a lower derivative theory.

People sometimes imagine that the energy of a higher derivative theory decays
with time. That is not true. Provided one is dealing with a complete system,
and provided there is no external time dependence, the energy of a higher
derivative system is conserved, just as it would be under those conditions
for a lower derivative theory. This conservation is apparent for our higher 
derivative oscillator from expression (\ref{H3}). 

The physical problem with nondegenerate higher derivative theories is not that
their energies decay to lower and lower values. The problem is rather that 
certain sectors of the theory become arbitrarily highly excited when one is 
dealing with an interacting, continuum field theory which has nondegenerate 
higher derivatives. To understand this I must digress to remind you of some 
familiar facts about the Hydrogen atom. 

If you consider Hydrogen in isolation, there is an infinite tower of stationary 
states. However, if you allow the Hydrogen atom to interact with 
electromagnetism only the ground state is stationary; all the excited states
decay through the emission of a photon. Why is this so? It certainly is {\it 
not} because ``the system wants to lower its energy.'' The energy of the full 
system is constant, the binding energy released by the decaying atom being 
compensated by the energy of the recoil photon. Yet the decay always takes 
place, and rather quickly. The reason is that decay is terrifically favored 
by entropy. If we prepare the Hydrogen atom in an excited state, with no 
photons present, there is {\it one} way for the atom to remain excited, 
whereas there are an {\it infinite} number of ways for it to decay because 
the recoil photon could go off in any direction.

Now consider an interacting, continuum field theory which possesses the
Ostrogradskian instability. In particular consider its likely particle
spectrum about some ``empty'' solution in which the field is constant.
Because the Hamiltonian is linear in all but one of the conjugate momenta
we can increase or decrease the energy by moving different directions in
phase space. Hence there must be both positive energy and negative energy
particles --- just as there are in our higher derivative oscillator. Just as 
in that point particle model, the same continuum field must carry the creation
and annihilation operators of {\it both} the positive and the negative energy 
particles. If the theory is interacting at all --- that is, if its Lagrangian 
contains a higher than quadratic power of the field --- then there will be 
interactions between positive and negative energy particles. Depending upon 
the interaction, the empty state can decay into some collection of positive 
and negative energy particles. The details don't really matter, all that 
matters is the counting: there is {\it one} way for the system to stay empty 
versus a continuous {\it infinity} of ways for it to decay. This infinity is 
even worse than for the Hydrogen atom because it includes not only all the 
directions that recoil particles of fixed energies could go but also the fact 
that the various energies can be arbitrarily large in magnitude provided they 
sum to zero. Because of that last freedom the decay is instantaneous. And the 
system doesn't just decay once! It is even {\it more} entropicly favored 
for there to be two decays, and better yet for three, etc. You can see 
that such a system instantly evaporates into a maelstrom of positive and 
negative energy particles. Some of my mathematically minded colleagues would 
say it isn't even defined. I prefer to simply observe that no theory of this 
kind can describe the universe we experience in which all particles have 
positive energy and empty space remains empty.

Note that we only reach this conclusion if the higher derivative theory 
possesses both interactions and continuum particles. Our point particle
oscillator has no interactions, so its negative energy degree of freedom is
harmless. Of course it is also completely unobservable! However, it is 
conceivable we could couple this higher derivative oscillator to a discrete 
system without engendering an instability. The feature that drives the 
instability when continuum particles are present is the vast entropy of
phase space. Without that it becomes an open question whether or not there is
anything wrong with a higher derivative theory. Of course we live in a 
continuum universe, and any degree of freedom we can observe must be
interacting, so these are very safe assumptions. However, people sometimes 
delude themselves that there is no problem with continuum, interacting higher 
derivative models of the universe on the basis of studying higher derivative 
systems which could never describe the universe because they either lack 
interactions or else continuum particles.

In this sub-section we have learned:
\begin{enumerate}
\item{The Ostrogradskian instability does not drive the dynamical variable to
a special, constant value but rather to a special kind of time dependence.}
\item{A dynamical variable which experiences the Ostrogradskian instability 
will carry both positive and negative energy creation and annihilation 
operators.}
\item{If the system interacts then the ``empty'' state can decay into a 
collection of positive and negative energy excitations.}
\item{If the system is a continuum field theory the vast entropy at infinite
momentum will make the decay instantaneous.}
\end{enumerate}

\subsection{Perturbation Theory}
\label{sub:3.2}

People sometimes mistakenly believe that the Ostrogradskian instability is
avoided if higher derivatives are segregated to appear only in interaction
terms. This is not correct if one considers the theory on a fundamental level.
One can see from the construction of section {\ref{sec:2} that the fact of
Ostrogradski's Hamiltonian being unbounded below depends only upon 
nondegeneracy, irrespective of how one organizes any approximation technique.
However, there is a way of imposing constraints to make the theory agree with
its perturbative development. If this is done then there are no more higher 
derivative degrees of freedom, however, one typically loses unitarity,
causality and Lorentz invariance on the nonperturbative level.

I constructed the higher derivative oscillator (\ref{HDO}) so that its
higher derivatives vanish when $g \!=\! 0$. If we solve the Euler-Lagrange 
equation (\ref{HDE}) exactly, without employing perturbation theory, there 
are four linearly independent solutions (\ref{gensol}) corresponding to a 
positive energy oscillator of frequency $k_+$ and a negative energy 
oscillator of frequency $k_-$. However, we might instead regard the parameter 
$g$ as a coupling constant and solve the equations perturbatively. This means
substituting the ansatz,
\begin{equation}
q_{\rm pert}(t) = \sum_{n=0}^{\infty} g^n x_n(t) \; , \label{pertan}
\end{equation}
into the Euler-Lagrange equation (\ref{HDE}) and segregating terms according
to powers of $g$. The resulting system of equations is,
\begin{eqnarray}
\ddot{x}_0 + \omega^2 x_0 & = & 0 \; , \label{E0} \\
\ddot{x}_1 + \omega^2 x_1 & = & -\frac1{\omega^2} x^{(4)}_0 \; , \\
\ddot{x}_2 + \omega^2 x_2 & = & -\frac1{\omega^2} x^{(4)}_1 \; , 
\end{eqnarray}
and so on. Because the zeroth order equation involves only second derivatives,
its solution depends upon only two pieces of initial value data,
\begin{equation}
x_0(t) = q_0 \cos(\omega t) + \frac{\dot{q}_0}{\omega} \sin(\omega t) \; .
\end{equation}
The first correction is,
\begin{equation}
x_1(t) = -\frac{\omega t}2 q_0 \sin(\omega t) + \frac{t}2 \dot{q}_0 
\cos(\omega t) - \frac1{2 \omega} \dot{q_0} \sin(\omega t) \; , 
\end{equation}
and it is easy to see that the sum of all corrections gives,
\begin{equation}
q_{\rm pert}(t) = q_0 \cos(k_+ t) + \frac{\dot{q}_0}{k_+} \sin(k_+ t) \; .
\label{pertsol}
\end{equation}

What is the relation of the perturbative solution (\ref{pertsol}) to the 
general one (\ref{gensol})? The perturbative solution is what results if
we change the theory by imposing the constraints,
\begin{eqnarray}
\ddot{q}(t) = - k_+^2 q(t) \qquad & \Longleftrightarrow & \qquad P_2 = 
\frac{m}2 \Bigl(1 \!-\! \sqrt{1 \!-\! 4g}\Bigr) Q_1 \; , \label{C1} \\
q^{(3)}(t) = - k_+^2 \dot{q}(t) \qquad & \Longleftrightarrow & \qquad P_1 =
\frac{m}2 \Bigl(1 \!+\! \sqrt{1 \!-\! 4g}\Bigr) Q_2 \; . \label{C2}
\end{eqnarray}
Under these constraints the Hamiltonian becomes,
\begin{equation}
H_{\rm pert} = \sqrt{1 \!-\! 4g} \Bigl( \frac{m}2 Q_2^2 + \frac{m k_+^2}2 Q_1^2
\Bigr) \; ,
\end{equation}
which is indeed that of a single harmonic oscillator. From the full theory,
perturbation theory has retained only the solution whose frequency is well
behaved for $g \rightarrow 0$,
\begin{equation}
k_+ = \omega \Bigl(1 + \frac12 g + \frac78 g^2 + O(g^3) \Bigr) \; . 
\label{lowk}
\end{equation}
It has discarded the solution whose frequency blows up as $g \rightarrow 0$,
\begin{equation}
k_- = \frac{\omega}{\sqrt{g}} \Bigl(1 -\frac12 g - \frac58 g^2 + O(g^3)\Bigr) 
\; . \label{highk}
\end{equation}

So what's wrong with this? In fact there is nothing wrong with the procedure
for our model. If the constraints (\ref{C1}-\ref{C2}) are imposed at one 
instant, they remain valid for all times as a consequence of the full equation
of motion. However, that is only because our model is free of interactions. 
Recall that this same feature means the positive and negative energy degrees 
of freedom exist in isolation of one another, and there is no decay to 
arbitrarily high excitation as there would be for an interacting, continuum 
field theory. 

When interactions are present it is more involved but still possible to impose 
constraints which change the theory so that only the lower derivative, 
perturbative solutions remain. The procedure was first worked out by Ja\'en,
Llosa and Molina \cite{JLM}, and later, independently, by Eliezer and me 
\cite{EW}. To understand its critical defect suppose we change the 
``interaction'' of our higher derivative oscillator from a quadratic term to a 
cubic one,
\begin{equation}
-\frac{g m}{2 \omega^4} \, \ddot{q}^2 \longrightarrow -\frac{g m}{6 \ell 
\omega^4} \, \ddot{q}^3 \; .
\end{equation}
Here $\ell$ is some constant with the dimensions of a length. As with the
quadratic interaction, the new equation of motion is fourth order,
\begin{equation}
-m \Biggl[ \frac{d^2}{dt^2} \Bigl(\frac{g \ddot{q}^2}{2 \ell \omega^4}\Bigr) +
\ddot{q} + \omega^2 q \Biggr] = 0 \; , 
\end{equation}
Its general solution depends upon four pieces of initial value data. However,
by isolating the highest derivative term of the free theory,
\begin{equation}
\ddot{q} =- \omega^2 q -\frac{d^2}{dt^2} \Bigl(\frac{g \ddot{q}^2}{2 \ell
\omega^4} \Bigr) \; , \label{rewrite}
\end{equation}
and then iteratively substituting (\ref{rewrite}), we can delay the 
appearance of higher derivatives on the right hand side to any desired order
in the coupling constant $g$. For example, two iterations frees the right
hand side of higher derivatives up to order $g^2$,
\begin{eqnarray}
\ddot{q} & = & -\omega^2 q -\frac{d^2}{dt^2} \Biggl\{ \frac{g}{2 \ell \omega^4}
\Biggl[ -\omega^2 q - \frac{d^2}{dt^2} \Bigl( \frac{g \ddot{q}^2}{2 \ell 
\omega^4}\Bigr)\Biggr]^2\Biggr\} \; , \\
& = & -\omega^2 q + \frac{g}{\ell} \Bigl( \omega^2 q^2 \!-\! \dot{q}^2\Bigr)
+ \frac{g^2}{2 \ell^2 \omega^4} \, q \frac{d^2}{dt^2} \Bigl( \ddot{q}^2\Bigr)
\nonumber \\
& & \hspace{.5cm} 
- \frac{g^2}{2 \ell^2 \omega^6} \frac{d^2}{dt^2} \Bigl[ q \frac{d^2}{dt^2}
\Bigl( \ddot{q}^2 \Bigr)\Bigr] - \frac{g^3}{8 \ell^3 \omega^{12}} \frac{d^2}{
dt^2} \Biggl[ \frac{d^2}{dt^2} \Bigl( \ddot{q}^2\Bigr) \Biggr]^2 \; .
\end{eqnarray}
This obviously becomes complicated fast! However, the lower derivative
terms at order $g^2$ are simple enough to give if I don't worry about the 
higher derivative remainder,
\begin{equation}
\ddot{q} = -\omega^2 q + \frac{g}{\ell} \Bigl( \omega^2 q^2 \!-\! \dot{q}^2
\Bigr) + \frac{g^2}{\ell^2} \Bigl( -6 \omega^2 q^3 \!+\! 14 q \dot{q}^2\Bigr)
+ O(g^3) \; .
\end{equation}
If we carry this out to infinite order, {\it and drop the infinite derivative
remainder}, the result is an equation of the traditional form,
\begin{equation}
\ddot{q} = f(q,\dot{q}) \; .
\end{equation}
The canonical version of this equation gives the first of the desired 
constraints. The second is obtained from the canonical version of its
time derivative.

The constrained system we have just described is consistent on the perturbative
level, but not beyond. It does not follow from the original, exact equation.
That would be no problem if we could define physics using perturbation theory, 
but we cannot. Perturbation theory does not converge for any known interacting,
continuum field theory in $3\!+\!1$ dimensions! The fact that the constraints 
are not consistent beyond perturbation theory means there is a nonperturbative
amplitude for the system to decay to the arbitrarily high excitation in the 
manner described in sub-section \ref{sub:3.1}. The fact that the constraints 
treat time derivatives differently than space derivatives also typically leads 
to a loss of causality and Lorentz invariance beyond perturbation theory.

A final comment concerns the limit of small coupling constant, i.e., $g 
\rightarrow 0$. One can see from the frequencies (\ref{lowk}-\ref{highk}) of
our higher derivative oscillator that the negative energy frequency diverges
for $g \rightarrow 0$. Disingenuous purveyors of higher derivative models
sometimes appeal to people's experience with {\it positive energy} modes by
arguing that, ``the $k_-$ mode approaches infinite frequency for small 
coupling so it must drop out.'' That is false! The argument is quite correct
for an infinite frequency {\it positive} energy mode in a stable theory. In
that case exciting the mode costs an infinite amount of energy which would 
have to be drawn from de-exciting finite frequency modes. However, a {\it 
negative} energy mode doesn't decouple as its frequency diverges. Rather it 
couples {\it more strongly} because taking its frequency to infinity opens up 
more and more ways to balance its negative energy by exciting finite frequency,
positive energy modes.

\subsection{Quantization}
\label{sub:3.3}

People sometimes imagine that quantization might stabilize a system against
the Ostrogradskian instability the same way that it does for the Hydrogen atom
coupled to electromagnetism. This is a failure to understand correspondence
limits. Conclusions drawn from classical physics survive quantization unless
they depend upon the system either being completely excluded from some region
of the canonical phase space or else inhabiting only a small region of it. For
example, the classical instability of the Hydrogen atom (when coupled to 
electromagnetism) derives from the fact that the purely Hydrogenic part of the 
energy,
\begin{equation}
E_{\rm Hyd} = \frac{\Vert \vec{p}\Vert^2}{2m} - \frac{e^2}{\Vert \vec{x}\Vert} 
\; .
\end{equation}
can be made arbitrarily negative by placing the electron close to the nucleus
at fixed momentum. Because this instability depends upon the system being
in a very small region of the canonical phase space, one might doubt that it
survives quantization, and explicit computation shows that it does not.

In contrast, the Ostrogradskian instability derives from the fact that 
$P_1 Q_2$ can be made arbitrarily negative by taking $P_1$ either very negative,
for positive $Q_2$, or else very positive, for negative $Q_2$. {\it This covers 
essentially half the classical phase space!} Further, the variables $Q_2$ and 
$P_1$ commute with one another in Ostrogradskian quantum mechanics. So there
is no reason to expect that the Ostrogradskian instability is unaffected by 
quantization.

\subsection{Unitarity vs. Instability}
\label{sub:3.4}

Particle physicists who quantize higher derivative theories don't typically
recognize a problem with the stability. They maintain that the problem with
higher derivatives is a breakdown of unitarity. In this sub-section I will 
again have recourse to the higher derivative oscillator (\ref{HDO}) to explain 
the connection between the two apparently unrelated problems.

Let us find the ``empty'' state wavefunction, $\Omega(Q_1,Q_2)$ that has the
minimum excitation in both the positive and negative energy degrees of 
freedom. The procedure for doing this is simple: first identify the positive
and negative energy lowering operators $\alpha_{\pm}$ and then solve the
equations,
\begin{equation}
\alpha_+ \vert \Omega \rangle = 0 = \alpha_- \vert \Omega \rangle \; .
\label{wavef}
\end{equation}
We can recognize the raising and lowering operators by simply expressing the 
general solution (\ref{gensol}) in terms of exponentials,
\begin{eqnarray}
\lefteqn{q(t) = \frac12 (A_+ \!+\! i B_+) e^{-ik_+ t} + \frac12 (A_+ \!-\! 
i B_+) e^{ik_+ t} } \nonumber \\
& & \hspace{2cm} + \frac12 (A_- \!+\! i B_-) e^{-ik_-t } + \frac12 (A_- \!-\! 
i B_-) e^{ik_- t} \; .
\end{eqnarray}
Recall that the $k_+$ mode carries positive energy, so its lowering operator
must be proportional to the $e^{-ik_+ t}$ term,
\begin{eqnarray}
\alpha_+ & \sim & A_+ + i B_+ \; , \\
& \sim & \frac{m k_+}2 \Bigl(1 \!+\! \sqrt{1 \!-\! 4g}\Bigr) Q_1 + i P_1
- k_+ P_2 - \frac{i m}2 \Bigl(1 \!-\! \sqrt{1 \!-\! 4g}\Bigr) Q_2 \; .
\end{eqnarray}
The $k_-$ mode carries negative energy, so its lowering operator must be 
proportional to the $e^{+i k_- t}$ term,
\begin{eqnarray}
\alpha_- & \sim & A_- - i B_- \; , \\
& \sim & \frac{m k_-}2 \Bigl(1 \!-\! \sqrt{1 \!-\! 4g}\Bigr) Q_1 - i P_1
- k_- P_2 + \frac{i m}2 \Bigl(1 \!+\! \sqrt{1 \!-\! 4g}\Bigr) Q_2 \; .
\end{eqnarray}
Writing $P_i = -i \frac{\partial}{\partial Q_i}$ we see that the unique 
solution to (\ref{wavef}) has the form,
\begin{equation}
\Omega(Q_1,Q_2) = N \exp\Biggl[-\frac{m \sqrt{1 \!-\! 4g}}{2 (k_+ \!+\! k_-)}
\Bigl(k_+ k_- Q_1^2 + Q_2^2\Bigr) - i \sqrt{g} m Q_1 Q_2\Biggr] \; . 
\label{true}
\end{equation}

The empty wave function (\ref{true}) is obviously normalizable, so it gives a
state of the quantum system. We can build a complete set of normalized 
stationary states by acting arbitrary numbers of $+$ and $-$ raising operators 
on it,
\begin{equation}
\vert N_+ , N_-\rangle \equiv \frac{(\alpha_+^{\dagger})^N_+}{\sqrt{N_+ !}}
\frac{(\alpha_-^{\dagger})^N_-}{\sqrt{N_- !}} \vert \Omega \rangle \; .
\end{equation}
On this space of states the Hamiltonian operator is unbounded below, just as
in the classical theory,
\begin{equation}
H \vert N_+ , N_- \rangle = \Bigl(N_+ k_+ - N_- k_-\Bigr) \vert N_+ , N_-
\rangle \; .
\end{equation}
This is the correct way to quantize a higher derivative theory. One evidence
of this fact is that classical negative energy states correspond to quantum
negative energy states as well.

Particle physicists don't quantize higher derivative theories as we just have.
What they do instead is to regard the negative energy lowering operator as a 
positive energy raising operator. So they define a ``ground state'' $\vert 
\overline{\Omega} \rangle$ which obeys the equations,
\begin{equation}
\alpha_+ \vert \overline{\Omega} \rangle = 0 = \alpha_-^{\dagger} \vert 
\overline{\Omega} \rangle \; . \label{falsewave}
\end{equation}
The unique wave function which solves these equations is,
\begin{equation}
\overline{\Omega}(Q_1,Q_2) = N \exp\Biggl[-\frac{m \sqrt{1 \!-\! 4g}}{2 (k_-
\!-\! k_+)} \Bigl(k_+ k_- Q_1^2 - Q_2^2\Bigr) + i \sqrt{g} m Q_1 Q_2 \Biggr]
\; . \label{wrong}
\end{equation}
This wave function is {\it not} normalizable, so it doesn't correspond to a
state of the quantum system. At this stage we should properly call a halt to
the analysis because we aren't doing quantum mechanics anymore. The 
Schrodinger equation $H \psi(Q) = E \psi(Q)$ is just a second order 
differential equation. It has two linearly independent solutions {\it for 
every} energy $E$: positive, negative, real, imaginary, quaternionic --- it 
doesn't matter. The thing that puts the ``quantum'' in quantum mechanics is 
requiring that the solution be normalizable. Many peculiar things can happen
if we abandon allow normalizability \cite{RPW1,TW0}.

However, my particle theory colleagues ignore this little problem and define 
a completely formal ``space of states'' based upon $\vert\overline{\Omega}
\rangle$,
\begin{equation}
\vert \overline{N_+ , N_-}\rangle \equiv \frac{(\alpha_+^{\dagger})^{N_+}}{
\sqrt{N_+ !}} \frac{(\alpha_-)^{N_-}}{\sqrt{N_- !}} \vert \overline{\Omega} 
\rangle \; .
\end{equation}
None of these wavefunctions is any more normalizable than $\overline{\Omega}(
Q_1,Q_2)$, so not a one of them corresponds to a state of the quantum system.
However, they are all positive energy eigenfunctions,
\begin{equation}
H \vert \overline{N_+ , N_-} \rangle = \Bigl(N_+ k_+ + N_- k_-\Bigr) \vert 
\overline{N_+ , N_-} \rangle \; .
\end{equation}
My particle physics colleagues typically say they {\it define} $\vert 
\overline{\Omega}\rangle$ to have unit norm. Because they have not changed the
commutation relations,
\begin{equation}
[\alpha_+,\alpha_+^{\dagger}] = 1 = [\alpha_-,\alpha_-^{\dagger}] \; , 
\end{equation}
the norm of any state with odd $N_-$ is negative! The lowest of these is,
\begin{equation}
\langle \overline{0,1} \vert \overline{0,1}\rangle =
\langle \overline{\Omega} \vert \alpha_-^{\dagger} \alpha_- \vert \overline{
\Omega} \rangle = - \langle \overline{\Omega} \vert \overline{\Omega} \rangle
\; .
\end{equation}
As I pointed out above, the reason this has happened is that we aren't
doing quantum mechanics any more. We ought to use the normalizable, but
indefinite energy eigenstates. What particle physicists do instead is to 
reason that because the probabilistic interpretation of quantum mechanics 
requires norms to be positive, the negative norm states must be excised from 
the space of states. At this stage good particle physicists note that that 
the resulting model fails to conserve probability \cite{KS}. Just as the 
correctly-quantized, indefinite-energy theory allows processes which mix 
positive and negative energy particles, so too the indefinite-norm theory 
allows processes which mix positive and negative norm particles. It only 
conserves probability on the space of ``states'' which includes both kinds of 
norms. If we excise the negative norm states then probability is no longer 
conserved.

So good particle physicists reach the correct conclusion --- that nondegenerate
higher derivative theories can't describe our universe --- by a somewhat
illegitimate line of reasoning. But who cares? They got the right answer! Of
course {\it bad} particle physicists regard the breakdown of unitarity as a 
challenge for inspired tinkering to avoid the problem. Favorite ploys are the 
Lee-Wick reformulation of quantum field theory \cite{LW} and nonperturbative 
resummations. The analysis also typically involves the false notion that 
high frequency ghosts decouple, which I debunked at the end of sub-section 
\ref{sub:3.2}. When the final effort is written up and presented to the world, 
some long-suffering higher derivative expert gets called away from his research 
to puzzle out what was done and explain why it isn't correct. {\it Sigh}. The 
problem is so much clearer in its negative energy incarnation! I could list 
many examples at this point, but I will confine myself to citing a full-blown 
paper debunking one of them \cite{TW1}. It is also appropriate to note that 
Hawking and Hertog have previously called attention to the mistake of 
quantizing higher derivative theories using nonnormalizable wave functions 
\cite{HH}.

\subsection{Constraints}
\label{sub:3.5}

The only way anyone has ever found to avoid the Ostrogradskian instability on a
nonperturbative level is by violating the single assumption needed to make 
Ostrogradski's construction: nondegeneracy. Higher derivative theories for 
which the definition of the highest conjugate momentum (\ref{nondeg}) cannot be
inverted to solve for the highest derivative can sometimes be stable. An 
interesting example of this kind is the rigid, relativistic particle studied by
Plyushchay \cite{MSP,DZ}.

Degeneracy is of great importance because {\it all theories which possess 
continuous symmetries are degenerate,} irrespective of whether or not they
possess higher derivatives. A familiar example is the relativistic point 
particle, whose dynamical variable is $X^{\mu}(\tau)$ and whose Lagrangian is,
\begin{equation}
L = -m \sqrt{-\eta_{\mu\nu} \dot{X}^{\mu} \dot{X}^{\nu}} \; .
\end{equation}
The conjugate momentum is,
\begin{equation}
P_{\mu} \equiv \frac{m \dot{X}_{\mu}}{\sqrt{-\dot{X}^2}} \; .
\end{equation}
Because the right hand side of this equation is homogeneous of degree zero
one can not solve for $\dot{X}^{\mu}$. The associated continuous symmetry is
invariance under reparameterizations $\tau \rightarrow \tau'(\tau)$,
\begin{equation}
X^{\mu}(\tau) \longrightarrow X^{\prime \mu}(\tau) \equiv X^{\mu}\Bigl({\tau'
}^{-1}(\tau)\Bigr) \; .
\end{equation}

The cure for symmetry-induced degeneracy is simply to fix the symmetry by
imposing gauge conditions. Then the gauge-fixed Lagrangian should no longer 
be degenerate in terms of the remaining variables. For example, we might 
parameterize so that $\tau = X^0(\tau)$, in which case the gauge-fixed
particle Lagrangian is,
\begin{equation}
L_{\rm GF} = -m \sqrt{1 - \dot{\vec{X}} \cdot \dot{\vec{X}} } \; .
\end{equation}
In this gauge the relation for the momenta is simple to invert,
\begin{equation}
P_i \equiv \frac{m \dot{X}_i}{\sqrt{1 - \dot{\vec{X}} \cdot \dot{\vec{X}} }} 
\qquad \Longleftrightarrow \qquad \dot{X}^i = \frac{P^i}{\sqrt{m^2 + \vec{P}
\cdot \vec{P}}} \; .
\end{equation}

When a continuous symmetry is used to eliminate a dynamical variable, the
equation of motion of this variable typically becomes a {\it constraint}. 
For symmetries enforced by means of a compensating field --- such as local 
Lorentz invariance is with the antisymmetric components of the vierbein
\cite{RPW2} --- the associated constraints are tautologies of the form $0 = 0$. 
Sometimes the constraints are nontrivial, but implied by the equations of 
motion. An example of this kind is the relativistic particle in our synchronous
gauge.  The equation of the gauge-fixed zero-component just tells us the 
Hamiltonian is conserved,
\begin{equation}
\frac{d}{d\tau} \Biggl( \frac{m \dot{X}_0}{\sqrt{-\eta_{\mu\nu} \dot{X}^{\mu} 
\dot{X}^{\nu}}} \Biggr) = 0 \longrightarrow \frac{d}{dt} \Bigl(
\sqrt{m^2 + \vec{p} \cdot \vec{p} } \Bigr) = 0 \; .
\end{equation}
And sometimes the constraints give nontrivial relations between the canonical 
variables that generate residual, time-independent symmetries. In this case
another degree of freedom can be removed (``gauge fixing counts twice,'' as
van Nieuwenhuizen puts it). An example of this kind of constraint is Gauss' 
Law in temporal gauge electrodynamics. 

Were it not for constraints of this last type, the analysis of a higher 
derivative 
theory with a gauge symmetry would be straightforward. One would simply fix 
the gauge and then check whether or not the gauge-fixed Lagrangian depends 
nondegenerately upon higher time derivatives. If it did, the conclusion would
be that the theory suffers the Ostrogradskian instability. However, when 
constraints of the third type are present one must check whether or not they
affect the instability. This is highly model dependent but a very simple rule
seems to be generally applicable: {\it if the number of gauge constraints is
less than the number of unstable directions in the canonical phase space then 
there is no chance for avoiding the problem}. Because the number of constraints
for any symmetry is fixed, whereas the number of unstable directions increases
with the number of higher derivatives, one consequence is that gauge 
constraints can at best avoid instability for some fixed number of higher
derivatives. For example, the constraints of the second derivative model of
Plyushchay are sufficient to stabilize the system \cite{MSP,DZ}, but one would
expect it to become unstable if third derivatives were added.

People sometimes make the mistake of believing that the Ostrogradskian 
instability can be avoided with just a single, global constraint on the 
Hamiltonian. For example, Boulware, Horowitz and Strominger \cite{BHS} showed 
the energy is zero for any asymptotically flat solution of the higher 
derivative field equations derived from the Lagrangian,
\begin{equation}
\mathcal{L} = \alpha R^2 \sqrt{-g} + \beta R^{\mu\nu} R_{\mu \nu} \sqrt{-g}
\; .
\end{equation}
As I explained in sub-section \ref{sub:3.1}, the nature of the Ostrogradskian 
instability is not that the energy decays but rather that the system evaporates
to a very highly excited state of compensating, positive and negative energy 
degrees of freedom. As long as $\beta \neq 0$, there are six independent, 
higher derivative momenta at each space point, whereas there are only four 
local constants --- or five if $\alpha$ and $\beta$ are such as to give local
conformal invariance. Hence there are two (or one) unconstrained instabilities 
per space point. There are an infinite number of space points, so the addition 
of a single, global constraint does not change anything. I should point out 
that Boulware, Horowitz and Strominger were aware of this, cf. their discussion
of the dipole instability. 

The case of $\beta = 0$ is special, and significant for the next section. If
$\alpha$ has the right sign that model has long been known to have positive
energy \cite{AAS0,AS}. This result in no way contradicts the previous analysis.
When $\beta = 0$ the terms which carry second derivatives are contracted in 
such way that only a single component of the metric carries higher derivatives.
So now the counting is {\it one} unstable direction per space point versus four
local constraints. Hence the constraints can win, and they do if $\alpha$ has 
the right sign.

\subsection{Nonlocality}
\label{sub:3.6}

I would like to close this section by commenting on the implications of 
Ostrogradski's theorem for fully nonlocal theories. In addition to nonlocal 
quantum field theories \cite{KW0,CHY,JJ} this is relevant to string field 
theory \cite{GJ1,GJ2,GJ3}, to noncommutative geometry \cite{NC,HPR}, to 
regularization techniques \cite{EMKW,KW1,KW2} and even to theories of cosmology 
\cite{TW2,SW1,BMS}. The issue in each case is whether or not we can think of 
the fully nonlocal theory as the limit of a sequence of ever higher derivative 
theories. When such a representation is possible the nonlocal theory must
inherit the Ostrogradskian instability.

The higher derivative representation is certainly valid for string field
theory because, otherwise, there would be cuts and poles that would interfere
with perturbative unitarity. So string field theory suffers from the
Ostrogradskian instability \cite{EW}. The same is true for theories where the 
nonlocality is of limited extent in time \cite{RPW3}, although not everyone
agrees \cite{JL,RPW4}. However, when the nonlocality involves inverse 
differential operators there need be no problem \cite{EW,SW1}. Indeed, the
effective action of any quantum field theory is nonlocal in this way 
\cite{BGVZ,BM}! Nor is there necessarily any problem when the nonlocality 
arises in the form of algebraic functions of local actions \cite{BNW}.

\section{$\Delta R[g] = f(R)$ Theories}
\label{sec:4}

From the lengthy argumentation of the previous two sections one might conclude
that the only potentially stable, local modification of gravity is a 
cosmological constant,
$\Delta R[g] = - 2\Lambda$. However, a close analysis of sub-section 
\ref{sub:3.5} reveals that it is also possible to consider algebraic functions
of the Ricci scalar. In this section I first explain why such theories can 
avoid the Ostrogradskian instability. I then demonstrate that they are 
equivalent to general relativity with a minimally coupled scalar, provided
we ignore matter. Finally, I exploit this equivalence, with the construction 
described in the Introduction, to show how $f(R)$ can be chosen to enforce
any evolution $a(t)$.

\subsection{Why They Can Be Stable}
\label{sub:4.1}

The alert reader will have noted that the $R + R^2$ model \cite{AAS0,AS}
avoids the Ostrogradskian instability. It does this by violating Ostrogradski's
assumption of nondegeneracy: the tensor indices of the second derivative terms
in the Ricci scalar are contracted together so that only a single component of
the metric carries higher derivatives. This component does acquire a new,
higher derivative degree of freedom, and the energy of this degree of freedom
is indeed opposite to that of the corresponding lower derivative degree of
freedom, just as required by Ostrogradski's analysis. However, that lower
derivative degree of freedom is the {\it Newtonian potential}. It carries
negative energy, but it is also completely fixed in terms of the other metric 
and matter fields by the $g_{00}$ constraint. So the only instability 
associated with it is gravitational collapse. Its higher derivative counterpart
has positive energy, at least on the kinetic level; it can still have a 
bad potential, and the model is indeed only stable for one sign of 
the $R^2$ term.

None of these features depended especially upon the higher derivative term 
being $R^2$. Any function for the Ricci scalar would work as well. Note that
we cannot allow derivatives of the Ricci scalar, because Ostrogradski's theorem
says the next higher derivative degree of freedom would carry negative energy
and there would be no additional constraints to protect it. We also cannot
permit more general contractions of the Riemann tensor because then other
components of the metric would carry higher derivatives. These components are 
positive energy in general relativity, so their higher derivative counterparts
would be negative, and there would again be no additional constraints to 
protect the theory against instability.

\subsection{Equivalent Scalar Representation}
\label{sub:4.2}

The general Lagrangian we wish to consider takes the form,
\begin{equation}
\mathcal{L} = \frac1{16 \pi G} \Bigl( R + f(R)\Bigr) \sqrt{-g} \; .
\end{equation}
If we ignore the coupling to matter the modified gravitational field equation
consists of the vanishing of the following tensor,
\begin{equation}
\frac{16 \pi G}{\sqrt{-g}} \frac{\delta S}{\delta g^{\mu\nu}} = [1 \!+\! f'(R)]
R_{\mu\nu} - \frac12 [R \!+\! f(R)] g_{\mu\nu} + g_{\mu\nu} [f'(R)]^{;\rho}_{
~\rho} - [f'(R)]_{;\mu\nu} \; . \label{MGR}
\end{equation}
There is an old procedure for reformulating this as general relativity with a 
minimally coupled scalar. I don't know whom to credit, but I will give the 
construction.

The first step is to define an ``equivalent'' theory with an auxiliary field 
$\phi$ which is defined by the relation.
\begin{equation}
\phi \equiv 1 + f'(R) \qquad \Longleftrightarrow \qquad R = \mathcal{R}(\phi)
\; .
\end{equation}
Inverting the relation determines the Ricci scalar as an algebraic function
of $\phi$. We can then define an auxiliary potential for $\phi$ by Legendre 
transformation,
\begin{equation}
U(\phi) \equiv \Bigl(\phi \!-\! 1\Bigr) \mathcal{R}(\phi) - 
f\Bigl(\mathcal{R}(\phi)\Bigr) \qquad \Longrightarrow \qquad U'(\phi) =
\mathcal{R}(\phi) \; .
\end{equation}
Now consider the equivalent scalar-tensor theory whose Lagrangian is,
\begin{equation}
\mathcal{L}_{\rm E} \equiv \frac1{16 \pi G} \Bigl(\phi R - U(\phi)\Bigr) 
\sqrt{-g} \; .
\end{equation}
Its field equations are,
\begin{eqnarray}
\frac{16 \pi G}{\sqrt{-g}} \frac{\delta S_{\rm E}}{\delta \phi} & = & R - 
U'(\phi) = 0 \; , \label{E11} \\
\frac{16 \pi G}{\sqrt{-g}} \frac{\delta S_{\rm E}}{\delta g^{\mu\nu}} & = & 
\phi R_{\mu\nu} - \frac12 \Bigl(\phi R \!-\! U(\phi)\Bigr) g_{\mu\nu} + 
g_{\mu\nu} \phi^{;\rho}_{~\rho} - \phi_{\mu\nu} = 0 \; . \label{E12}
\end{eqnarray}
The scalar equation (\ref{E11}) implies $\phi \!=\! 1 \!+\! f'(R)$, whereupon 
the tensor equations (\ref{E12}) reproduce the original modified gravity 
equations (\ref{MGR}).

The final step is to define a new metric $\widetilde{g}_{\mu\nu}$ and a new
scalar $\varphi$ by the change of variables,
\begin{eqnarray}
\widetilde{g}_{\mu\nu} \equiv \phi \, g_{\mu\nu} \qquad & \Longleftrightarrow &
\qquad g_{\mu\nu} = \exp\Bigl[-\sqrt{\frac{4\pi G}3} \, \varphi\Bigr] \,
\widetilde{g}_{\mu\nu} \; , \label{gtrans} \\
\varphi \equiv \sqrt{\frac3{4\pi G}} \, \ln(\phi) \qquad & \Longleftrightarrow 
& \qquad \phi = \exp\Bigl[\sqrt{\frac{4\pi G}3} \, \varphi\Bigr] \; .
\end{eqnarray}
In terms of these variables the equivalent Lagrangian takes the form,
\begin{equation}
\mathcal{L}_E = \frac1{16 \pi G} \widetilde{R} \sqrt{-\widetilde{g}}
-\frac12 \partial_{\mu} \varphi \partial_{\nu} \varphi \,
\widetilde{g}^{\mu\nu} \sqrt{-\widetilde{g}} - V(\varphi) 
\sqrt{-\widetilde{g}} \; , \label{finalL}
\end{equation}
where the scalar potential is,
\begin{equation}
V(\varphi) \equiv \frac1{16 \pi G} U\Biggl(\exp\Bigl[
\sqrt{\frac{4 \pi G}3} \, \varphi\Bigr]\Biggr) \exp\Bigl[-\sqrt{\frac{16 \pi 
G}3} \, \varphi\Bigr] \; .
\end{equation}
This is general relativity with a minimally coupled scalar, as claimed.

\subsection{Reconstructing $f(R)$ from Cosmology}
\label{sub:4.3}

I want to show how to choose $f(R)$ to support an arbitrary $a(t)$.\footnote{
For a somewhat different construction which achieves the same end, see
\cite{NO0,NO00}.} Recall from the Introduction that one can choose the 
potential of a quintessence model such as (\ref{finalL}) to support any 
homogeneous and isotropic cosmology for its metric $\widetilde{g}_{\mu\nu}$. 
However, we cannot immediately exploit this construction because it is the 
metric $g_{\mu\nu}$ which is assumed known, not $\widetilde{g}_{\mu\nu}$. We 
must explain how to infer the one from the other without knowing $f(R)$.

Because the relation (\ref{gtrans}) between $g_{\mu\nu}$ and $\widetilde{g}_{
\mu\nu}$ is a conformal transformation, it makes sense to work in a coordinate
system in which each metric is conformal to flat space. This is accomplished
by changing from co-moving time $t$ to conformal time $\eta$ though the 
relation, $d\eta = dt/a(t)$,
\begin{equation}
ds^2 = -dt^2 + a^2(t) d\vec{x} \cdot d\vec{x} = a^2 \Bigl(-d\eta^2 + d\vec{x}
\cdot d\vec{x}\Bigr) \; .
\end{equation}
The $\widetilde{g}_{\mu\nu}$ element takes the same form in conformal 
coordinates, but note that its different scale factor implies a different
co-moving time,
\begin{equation}
d\widetilde{s}^2 = \widetilde{a}^2 \Bigl(-d\eta^2 + d\vec{x} \cdot 
d\vec{x}\Bigr) = -d\widetilde{t}^{~2} + \widetilde{a}^2(\widetilde{t}\,) 
d\vec{x} \cdot d\vec{x} \; .
\end{equation}
From relation (\ref{gtrans}) we infer,
\begin{equation}
a(t) = \widetilde{a}(\widetilde{t}\,) \exp\Bigl[-\sqrt{\frac{\pi G}3} \, 
\varphi_0(\widetilde{t}\,) \Bigr] \; . \label{keyrel}
\end{equation}

We denote differentiation with respect to $\eta$ by a prime, and one should
note the relation between derivatives with respect to the various times,
\begin{equation}
\frac{\partial}{\partial \eta} = a \frac{\partial}{\partial t} = 
\widetilde{a} \frac{\partial}{\partial \widetilde{t}} \; .
\end{equation}
Differentiating the logarithm of (\ref{keyrel}) with respect to $\eta$ and 
using the relation (\ref{twoeqns}) between $\widetilde{a}$ and $\varphi_0$ 
gives,
\begin{equation}
\frac{a'}{a} = \frac{\widetilde{a}'}{\widetilde{a}} 
-\sqrt{\frac{\pi G}3} \, \varphi_0' 
= \frac{\widetilde{a}'}{\widetilde{a}} -\sqrt{-\frac1{12} \widetilde{a}'} \; .
\end{equation}
This is a nonlinear but first order differential equation for the variable
$\widetilde{a}$ in terms of the known function, $a(t(\eta))$. At the worst
it can be solved numerically.

Once we have $\widetilde{a}$ the potential $V(\varphi)$ can be constructed
using the procedure explained in the Introduction. We then compute the
auxiliary potential,
\begin{equation}
U(\phi) = 16 \pi G \phi^2 V\Bigl( \sqrt{\frac3{4\pi G}} \, \ln(\phi)\Bigr) \; .
\end{equation}
The auxiliary field can be expressed in terms of the Ricci scalar from the
algebraic relation,
\begin{equation}
U'(\phi) = R \qquad \Longleftrightarrow \qquad \phi = \Phi(R) \; .
\end{equation}
And we finally recover the function $f(R)$ by Legrendre transformation,
\begin{equation}
f(R) = \Bigl(\Phi(R) \!-\! 1\Bigr) R - U\Bigl(\Phi(R)\Bigr) \; .
\end{equation}

\section{Problems with $f(R) = -\frac{\mu^4}{R}$}
\label{sec:5}

In view of the construction of sub-section \ref{sub:4.3} it is not surprising
but rather {\it inevitable} that an $f(R)$ can be found to support late time
acceleration, or indeed, any other evolution. However, the method is not
guaranteed to produce a simple model, so the discovery that $f(R) = -\mu^4/R$
works is quite noteworthy \cite{CDTT,CCT}.\footnote{Although extensions 
involving $R^{\mu\nu} R_{\mu\nu}$ and $R^{\rho\sigma\mu\nu} 
R_{\rho\sigma\mu\nu}$ have also been studied \cite{CDDETT}, they must be ruled
out on account of the Ostrogradskian instability.} It may also be significant 
that models of this type seem to follow from fundamental theory \cite{NO1}.

To derive acceleration in this model consider its field equations,
\begin{equation}
\Bigl(1 \!+\! \frac{\mu^4}{R^2}\Bigr) R_{\mu\nu} - \frac12 \Bigl(1 \!-\!
\frac{\mu^4}{R^2}\Bigr) R g_{\mu\nu} + \Bigl(g_{\mu\nu} \square - D_{\mu}
D_{\nu}\Bigr) \frac{\mu^4}{R^2} = 8 \pi G T_{\mu\nu} \; . \label{theeqn}
\end{equation}
Setting $T_{\mu\nu} \!=\! 0$ and searching for constant Ricci scalar solutions
gives,
\begin{equation}
\Bigl(1 \!+\! \frac{\mu^4}{R^2}\Bigr) R_{\mu\nu} - \frac12 \Bigl(1 \!-\! 
\frac{\mu^4}{R^2}\Bigr) R g_{\mu\nu} = 0 \qquad \Longleftrightarrow \qquad
R_{\mu\nu} = \pm \frac{\sqrt{3}}4 \mu^2 g_{\mu\nu} \; .
\end{equation}
The plus sign corresponds to acceleration.

In addition to proposing the model, Carroll, Duvvuri, Trodden and Turner
\cite{CDTT} also showed that it suffers from a very weak tachyonic instability
in the absence of matter. Because the only new higher derivative degree of 
freedom resides in the Ricci scalar, we may as well derive an equation for it
alone from the trace of (\ref{theeqn}),
\begin{equation}
-R + \frac{3\mu^4}{R} + \square \Bigl(\frac{3\mu^4}{R^2}\Bigr) = 0 \; .
\end{equation}
Now perturb about the accelerated solution,
\begin{equation}
R = +\sqrt{3} \mu^2 + \delta R \quad \Longrightarrow \quad -2 \delta R
-\frac{2}{\sqrt{3} \mu^2} \square \delta R + O(\delta R^2) = 0 \; .
\end{equation}
By comparing the linearized equation for $\delta R$ with that of a positive
mass-squared scalar,
\begin{equation}
(\square - m^2) \varphi = 0 \; , \label{comp}
\end{equation}
we see that $\delta R$ behaves like a tachyon with $m^2 = -\sqrt{3} \mu^2$.
However, because explaining the current phase of acceleration requires
$\mu \sim 10^{-33}~{\rm eV}$, the resulting instability is not very serious.
I should note that the existence of a tachyonic instability in no way 
contradicts the Ostrogradskian analysis that this model's higher derivative
degree of freedom carries positive kinetic energy.

\subsection{Inside Matter}
\label{sub:5.1}

Dolgov and Kawasaki \cite{DK} showed that a radically different result emerges
when this model is considered inside a static distribution of matter,
\begin{equation}
T_{\mu\nu} = \rho \delta_{\mu}^0 \delta_{\nu}^0 \qquad {\rm with} \qquad 
8 \pi G \rho \equiv M^2 \gg \mu^2 \; . \label{Dolgov}
\end{equation}
In that case the trace of (\ref{theeqn}) gives,
\begin{equation}
-R + \frac{3 \mu^4}{R} + \square \Bigl(\frac{3 \mu^4}{R^2} \Bigr) = - M^2 \; .
\end{equation}
As might be expected, the static Ricci scalar solution in this case is 
dominated by $M$ rather than $\mu$,
\begin{equation}
R_0 = \frac12 \Bigl(M^2 \!+\! \sqrt{M^4 \!+\! 12 \mu^4}\Bigr) \simeq M^2 \; .
\end{equation}
Perturbing about this solution gives,
\begin{equation}
R = R_0 + \delta R \quad \Longrightarrow \quad - \delta R -\frac{3\mu^4}{
R_0^2} \delta R -\frac{6 \mu^4}{R_0^3} \square \delta R + O(\delta R^2) = 0\; .
\end{equation}
Comparing with the reference scalar (\ref{comp}) now reveals an enormous
tachyonic mass,
\begin{equation}
m^2 = -\frac{R_0}2 -\frac{R_0^3}{6 \mu^4} \simeq -\frac{M^6}{6 \mu^4} \; !
\end{equation}
Plugging in the numbers for the density of water ($\rho \sim 10^3~{\rm kg/m}^3$)
gives $M \sim 10^{-18}~{\rm eV}$, implying a tachyonic mass of magnitude
$\vert m\vert \sim 10^{12}~{\rm eV} = 10^3~{\rm GeV}$!

As disastrous as this problem might seem, Dick \cite{RD} and Nojiri and 
Odintsov \cite{NO2} have shown that it can be avoided by changing the model 
slightly,
\begin{equation}
f(R) = -\frac{\mu^4}{R} + \frac{\alpha}{2 \mu^2} R^2 \quad \Longrightarrow
\quad -R + \frac{3 \mu^4}{R} + 3 \square \Bigl( \frac{\mu^4}{R^2} + 
\frac{\alpha}{\mu^2} R \Bigr) = 0 \; . \label{Ext}
\end{equation}
Because an $R^2$ term has global conformal invariance, it makes no contribution
to the trace for constant $R$. Hence the cosmological solution of $R = +
\sqrt{3} \mu^2$ is not affected, nor is the static solution inside the matter
distribution (\ref{Dolgov}). However, the equation for linearized perturbations
inside matter changes to,
\begin{equation}
-\delta R - \frac{3 \mu^4}{R_0^2} \delta R + 3 \Bigl(-\frac{2 \mu^4}{R_0^3}
\!+\! \frac{\alpha}{\mu^2} \Bigr) \square \delta R = 0 \; .
\end{equation}
The instability of Dolgov and Kawasaki was driven by the smallness of $2\mu^4/
R_0^3$. By simply taking $\alpha$ positive and of order one the tachyon becomes
a positive mass-squared particle of $m^2 \sim \mu^2/\alpha$.

\subsection{Outside Matter}
\label{sub:5.2}

Marc Soussa and I analyzed force of gravity outside a matter distribution
\cite{SW2}. Although our analysis was for the original $f(R)= -\mu^4/R$
model, there would be only slight differences for the extended model 
(\ref{Ext}). So our result seems to foreclose this possibility, but see 
\cite{NO3}.

The tachyonic instability could be studied using the perturbed Ricci scalar,
but the gravitational force requires use of the metric. We perturbed about the 
de Sitter solution with Hubble constant $H = \mu/(48)^{\frac14}$ in co-moving 
coordinates,
\begin{equation}
ds^2 = -(1 \!-\! h_{00}) dt^2 + 2 a(t) h_{0i} dt dx^i + a^2(t) (\delta_{ij}
\!+\! h_{ij}) dx^i dx^j \quad {\rm with} \quad a(t) = e^{H t} \; .
\end{equation}
In the gauge,
\begin{equation}
h_{\mu\nu}^{~~,\nu} - \frac12 h_{\mu} + 3 h_{\mu}^{~\nu} [\ln(a)]_{,\nu} = 0 
\; ,
\end{equation}
with $h \equiv -h_{00} \!+\! h_{ii}$, the perturbed Ricci scalar takes the 
form,
\begin{equation}
\delta R = -\frac12 \partial^2 h + 2 H \partial_0 h \; . \label{deltaR}
\end{equation}
Our strategy was first to solve the de Sitter invariant equation for the
perturbed Ricci scalar, then reconstruct the gauge-fixed metric.

We assumed a matter density of the form,
\begin{equation}
\rho(t,\vec{x}) = \frac{3 M}{4 \pi R_g^3} \theta\Bigl(R_g - a(t) \vert \vec{x}
\vert \Bigr) \; .
\end{equation}
The exterior field equation has a simple expression in terms of the coordinate
$y \equiv a(t) H \vert \vec{x}\vert$,
\begin{equation}
\Biggl[\Bigl(1 \!-\! y^2\Bigr) \frac{d^2}{dy^2} + \frac2{y} \Bigl(1 \!-\! 2 y^2
\Bigr) \frac{d}{dy} + 12 \Biggr] \delta R = 0 \; .
\end{equation}
The solution takes the form,
\begin{equation}
\delta R = \beta_1 f_0(y) + \beta_2 f_{-1}(y) \; , \label{genform}
\end{equation}
where $f_0$ and $f_{-1}$ are hypergeometric functions whose series expansions
are,
\begin{eqnarray}
f_0(y) & = & 1 - 2 y^2 + \frac15 y^4 + \ldots \; , \\
f_{-1}(y) & = & \frac1{y} \Bigl( 1 - 7 y^2 + \frac{14}3 y^4 + \ldots \Bigr) 
\; .
\end{eqnarray}
We only need the behavior for small $y$ because $y \!=\! 1$ is the Hubble
radius! Matching to the source at $y = H R_g$ determines the combination 
coefficients to be,
\begin{equation}
\beta_1 \simeq \frac{3 G M}{R_g^3} \qquad , \qquad \beta_2 \simeq -12 G M H^3
\; .
\end{equation}
This last step might seem bogus because we needed to regard the mass density as
a small perturbation on the cosmological energy density $\mu^4$, whereas the
opposite would be the case for galaxies or clusters of galaxies. However, this
will only make changes of order one in the $\beta_i$'s. In particular, the
asymptotic solution must still take the form (\ref{genform}).

The next step is solving for the trace of the perturbed metric. It turns out
that relation (\ref{deltaR}) can also be expressed very simply using the
variable $y$,
\begin{equation}
\Biggl[ \Bigl(y^2 \!-\! 1\Bigr) \frac{d}{dy} + \frac1{y} \Bigl(5 y^2 \!-\!2
\Bigr) \Biggr] h'(y) = \frac{2}{H^2} \delta R \; .
\end{equation}
We only need to solve for the derivative of $h$ because that is what gives
the gravitational force in the geodesic equation. The solution is,
\begin{equation}
h'(y) = -\frac{2 G M}{H^2 R_g^3} y + O(y^3) \; .
\end{equation}
This should be compared to the general relativistic prediction,
\begin{equation}
h'_{\rm GR}(y) = -\frac{4 G M H}{y^2} + O(1) \qquad \Longrightarrow \qquad
\frac{h'}{h'_{\rm GR}} = \frac12 \Bigl(\frac{\Vert \vec{x}\Vert}{R_g}\Bigr)^3
\; .
\end{equation}
One consequence is that the force between the Milky Way and Andromeda galaxies
would be about a million times larger than predicted by general relativity!

\section{Conclusions}
\label{sec:6}

The potential of a quintessence scalar can be chosen to support any cosmology, 
but the epicyclic nature of this construction suggests we consider modifications
of gravity. Ostrogradski's theorem \cite{MO} limits local modifications of 
gravity to just algebraic functions of the Ricci scalar. Models of this form 
can give a late phase of cosmic acceleration such as we are currently 
experiencing. However, they can be tuned to give anything else as well. They 
seem every bit as epicyclic as scalar quintessence. Further, the $f(r) = 
-\mu^4/R$ model is problematic, both inside and outside matter 
sources.\footnote{Observations also rule out the somewhat different version 
of this model that results from regarding the connection and the metric
as independent, fundamental variables in the Palatini formalism \cite{AEMM}.}

An interesting and largely overlooked possibility for modifying gravity is 
the fully nonlocal effective action that results from quantum gravitational 
corrections. In weak field perturbation theory it has long been known that 
the most cosmologically significant one loop corrections are not of the $R^2$ 
form usually studied but rather of the form $R \ln(\square) R$ \cite{EMV}.
More potentially interesting is the possibility of very strong infrared
effects from the epoch of primordial inflation \cite{TW3,RBM}. 

It can be shown that quantum gravitational corrections to the inflationary 
expansion rate grow with time like powers of $\ln(a)$. Although suppressed by
very small coupling constants, the exponential growth in $a(t)$ during 
inflation must eventually cause the effect to become nonperturbatively strong 
\cite{TW4,TW5}. Similar secular growth occurs as well for minimally coupled 
scalar field theories \cite{OW1,OW2}, in which context Starobinski\u{\i} has 
developed a technique for summing the leading powers of $\ln(a)$ at each 
loop order \cite{AAS,SY}. If Starobinski\u{\i}'s technique can be generalized 
to quantum gravity \cite{RPW5,TW6} it might result in a nonlocal effective 
gravity theory for late time cosmology in which a large, bare cosmological 
constant is almost completely screened by a nonperturbative quantum 
gravitational effect. In such a formalism the current phase of acceleration 
might result from a very slight mismatch between the bare cosmological 
constant and the quantum effect which screens it. It is even conceivable that 
one could reproduce the phenomenological successes of MOND \cite{MM,SM} with 
such a nonlocal metric theory, although it would have to unstable against 
decay into galaxy-scale gravitational waves \cite{SW3}.

\vskip 1cm
\centerline{Acknowledgements}
It is a pleasure to acknowledge conversations and correspondence on this 
subject with S. Deser, A.D. Dolgov, D.A. Eliezer, S. Odintsov, M.E. Soussa,
A. Strominger and M. Trodden. This work was partially supported by NSF grant 
PHY-244714 and by the Institute for Fundamental Theory at the University of 
Florida.

\printindex

\begin{thebibliography}{99.}

\bibitem{BBN} R.H. Cyburt, B.D. Fields, K.A. Olive: Phys. Lett. \textbf{B567},
227 (2003), astro-ph/0302431

\bibitem{CDM} G. Bertone, D. Hooper, J. Silk: Phys. Rept. \textbf{405}, 279
(2005), hep-ph/0404175

\bibitem{MM} M. Milgrom: Ap. J. \textbf{270}, 365 (1983)

\bibitem{SM} R.H. Sanders and S.S. McGaugh: Ann. Rev. Astrophys \textbf{40}, 
263 (2002), astro-ph/0204521

\bibitem{GSKVK} G. Gentile, P. Salucci, U. Klein, D. Vergani, P. Kalberla:
Mon. Not. Roy. Astron. Soc. \textbf{351}, 903 (2004), astro-ph/0403154

\bibitem{JDB} J.D. Bekenstein: Phys. Rev. \textbf{D70}, 083509 (2004),
astro-ph/0403694

\bibitem{SMFB} C. Skordis, D.F. Mota, P.G. Ferreira, C. Boehm: Phys. Rev. Lett.
\textbf{96}, 011301 (2006), astro-ph/0505519

\bibitem{SNCP} R.A. Knop et al: Astrophys. J. \textbf{598}, 102 (2003),
astro-ph/0309368

\bibitem{SNST} A.D. Reiss et al: Astrophys. J. \textbf{607}, 665 (2004),
astro-ph/0402512

\bibitem{SNLS} P. Astier et al: astro-ph/0510447.

\bibitem{CMB} D. Spergel et al: Astrophys. J. Suppl. \textbf{148}, 175 (2003),
astro-ph/0302209

\bibitem{LSS} M. Tegmark et al: Phys. Rev. \textbf{D69}, 103501 (2004),
astro-ph/0310723

\bibitem{CW} C. Wetterich: Nucl. Phys. \textbf{B302}, 668 (1988)

\bibitem{PR} B. Ratra, P.J.E. Peebles: Phys. Rev. \textbf{D37}, 3406 (1988)

\bibitem{TW2} N.C. Tsamis, R.P. Woodard: Ann. Phys. \textbf{267}, 145 (1998),
hep-th/9712331

\bibitem{SRSS} T.D. Saini, S. Raychaudhury, V. Saini, A.A. Starobinski\u{\i}:
Phys. Rev. Lett. \textbf{85}, 1162 (2000), astro-ph/9910231

\bibitem{NO0} S. Capozziello, S. Nojiri, S.D. Odintsov: hep-th/0512118

\bibitem{MO} M. Ostrogradski: Mem. Ac. St. Petersbourg \textbf{VI 4}, 385
(1850)

\bibitem{JLM} X. Ja\'en, J. Llosa, A. Molina: Phys. Rev. \textbf{D34}, 2302
(1986)

\bibitem{EW} D.A. Eliezer, R.P. Woodard: Nucl. Phys. \textbf{B325}, 389
(1989)

\bibitem{RPW1} R.P. Woodard: Class. Quant. Grav. \textbf{10}, 483 (1993)

\bibitem{TW0} N.C. Tsamis, R.P. Woodard: Phys. Rev. \textbf{D36}, 3641 (1987)

\bibitem{KS} K.S. Stelle: Phys. Rev. \textbf{D16}, 953 (1977).

\bibitem{LW} T.D. Lee, G.C. Wick: Phys. Rev. \textbf{D2}, 1033 (1970)

\bibitem{TW1} N.C. Tsamis, R.P. Woodard: Ann. Phys. \textbf{168}, 457 (1986)

\bibitem{HH} S.W. Hawking, T. Hertog: Phys. Rev. \textbf{D65}, 103515 
(2002), hep-th/0107088

\bibitem{MSP} M.S. Plyushchay: Mod. Phys. Lett. \textbf{A4}, 837 (1989)

\bibitem{DZ} D. Zoller: Phys. Rev. Lett. \textbf{65}, 2236 (1990)

\bibitem{RPW2} R.P. Woodard: Phys. Lett. \textbf{B148}, 440 (1984)

\bibitem{BHS} D.G. Boulware, G.T. Horowitz, A. Strominger: Phys. Rev. Lett. 
\textbf{50}, 1726 (1983)

\bibitem{AAS0} A.A. Starobinski\u{\i}: Phys. Lett. \textbf{B91}, 99 (1980)

\bibitem{AS} A. Strominger: Phys. Rev. \textbf{D30}, 2257 (1984)

\bibitem{KW0} G. Kleppe, R.P. Woodard: Nucl. Phys. \textbf{B388}, 81 (1992)

\bibitem{CHY} T.C. Cheng, P.M. Ho, M.C. Yeh: Nucl. Phys. \textbf{B625}, 151
(2002), hep-th/0111160

\bibitem{JJ} A Jain, S.D. Joglekar: Int. J. Mod. Phys. \textbf{A19}, 3409
(2004), hep-th/0307208

\bibitem{GJ1} D.J. Gross, A. Jevicki: Nucl. Phys. \textbf{B283}, 1 (1987)

\bibitem{GJ2} D.J. Gross, A. Jevicki: Nucl. Phys. \textbf{B287}, 225 (1987)

\bibitem{GJ3} D.J. Gross, A. Jevicki: Nucl. Phys. \textbf{B293}, 29 (1987)

\bibitem{NC} A. Konechny, A. Schwarz: Phys. Rept. \textbf{360}, 353 (2002),
hep-th0107251

\bibitem{HPR} J.L. Hewett, F.J. Petriello, T.G. Rizzo: Phys. Rev. \textbf{D64},
075012 (2001), hep-ph/0010354

\bibitem{EMKW} D. Evens, J.W. Moffat, G. Kleppe, R.P. Woodard: Phys. Rev. 
\textbf{D43}, 499 (1991)

\bibitem{KW1} G. Kleppe, R.P. Woodard: Phys. Lett. \textbf{B253}, 331 (1991)

\bibitem{KW2} G. Kleppe, R.P. Woodard: Ann. Phys. \textbf{B221}, 106 (1993)

\bibitem{SW1} M.E. Soussa, R.P. Woodard: Class. Quant. Grav. \textbf{20},
2737 (2003), astro-ph/0302030

\bibitem{BMS} T. Biswas, A. Mazumdar, W. Siegel: hep-th/0508194

\bibitem{RPW3} R.P. Woodard: Phys. Rev. \textbf{A62}, 052105 (2000),
hep-th/0006207

\bibitem{JL} J. Llosa: hep-th/0201087

\bibitem{RPW4} R.P. Woodard: Phys. Rev. \textbf{A67}, 016102 (2003), 
hep-th/0207191

\bibitem{BGVZ} A.O.Barvinsky, Y.V. Gusev, G.A. Vilkovisky, V.V. Zhytnikov:
J. Math. Phys. \textbf{35}, 3525 (1994), gr-qc/9404061

\bibitem{BM} A.O. Barvinsky, V.F. Mukhanov: Phys. Rev. \textbf{D66}, 065007 
(2202), hep-th/0203132

\bibitem{BNW} D.L. Bennett, H.B. Nielsen, R.P. Woodard: Phys. Rev. 
\textbf{D57}, 1167 (1998, hep-th/9707088

\bibitem{NO00} S. Nojiri and S.D. Odintsov, hep-th/0601213

\bibitem{CDTT} S.M. Carroll, V. Duvvuri, M. Trodden, M.S. Turner: Phys. Rev.
\textbf{D70}, 043528 (2004), astro-ph/0306438

\bibitem{CCT} S. Capozziello, S. Carloni, A. Troisi: astro-ph/0306438

\bibitem{CDDETT} S.M. Carroll, A. De Felice, V. Duvvuri, D.A. Easson, 
M. Trodden, M.S. Turner: Phys. Rev. \textbf{D71}, 063513 (2005), 
astro-ph/0410031

\bibitem{NO1} S. Nojiri, S.D. Odintsov: Phys. Lett. \textbf{B576}, 5 (2003),
hep-th/0307071

\bibitem{DK} A.D. Dolgov, M. Kawasaki: Phys. Let. \textbf{B573}, 1 (2003),
astro-ph/0307285

\bibitem{RD} R. Dick: Gen. Rel. Grav. \textbf{36}, 217 (2004), gr-qc/0307052

\bibitem{NO2} S. Nojiri, S.D. Odintsov: Phys. Rev. \textbf{D68}, 123512 (2003),
hep-th/0307288.

\bibitem{SW2} M.E. Soussa, R.P. Woodard: Gen. Rel. Grav. \textbf{36}, 855 
(2004), astro-ph/0308114

\bibitem{NO3} S. Nojiri, S.D. Odintsov: Gen. Rel. Grav. \textbf{36}, 1765 
(2004), hep-th/0308176.

\bibitem{AEMM} M. Amarzguioui, \O. Elgaroy, D.F. Mota, T. Multam\"aki:
astro-ph/0510519

\bibitem{EMV} D. Espriu, T. Multam\"aki, E.C. Vagenas: Phys. Lett. 
\textbf{B628}, 197 (2005), gr-qc/0503033

\bibitem{TW3} N.C. Tsamis, R.P. Woodard: Ann. Phys. \textbf{238}, 1 (1995)

\bibitem{RBM} P. Martineau, R. Brandenberger: astro-ph/0510523

\bibitem{TW4} N.C. Tsamis, R.P. Woodard: Nucl. Phys. \textbf{B474}, 235 (1996),
hep-ph/9602315

\bibitem{TW5} N.C. Tsamis, R.P. Woodard: Ann. Phys. \textbf{253}, 1 (1997),
hep-ph/9602317

\bibitem{OW1} V.K. Onemli, R.P. Woodard: Class. Quant. Grav. \textbf{19},
4607 (2002), gr-qc/0204065

\bibitem{OW2} V.K. Onemli, R.P. Woodard: Phys. Rev. \textbf{D70}, 107301
(2004), gr-qc/0406098

\bibitem{AAS} A. A. Starobinski\u{\i}: Stochastic de Sitter (inflationary)
stage in the early universe. In: \textit{Field Theory, Quantum Gravity and
Strings}, ed by H.J. de Vega, N. Sanchez (Springer-Verlag, Berlin, 1986)
pp 107--126

\bibitem{SY} A.A. Starobinski\u{\i}, J.Yokoyama: Phys. Rev. \textbf{D50},
6357 (1994), astro-ph/\-9407016

\bibitem{RPW5} R.P. Woodard: Nucl. Phys. Proc. Suppl. \textbf{148}, 108 
(2005), astro-ph/0502556

\bibitem{TW6} N.C. Tsamis, R.P. Woodard: Nucl. Phys. \textbf{B724}, 295
(2005), gr-qc/0505115

\bibitem{SW3} M.E. Soussa, R.P. Woodard: Phys. Lett. \textbf{B578}, 253
(2004), astro-ph/0307358

\end{thebibliography}
\end{document}